\documentclass[11pt,a4paper]{article}
\usepackage[hyperref]{acl2021}
\usepackage{times}
\usepackage{latexsym}

\aclfinalcopy
\usepackage{helvet}
\usepackage{courier}
\usepackage{graphicx}
\usepackage{caption}
\usepackage{booktabs}
\usepackage{amsfonts}
\usepackage{xcolor,colortbl}
\usepackage{algpseudocode}
\usepackage{amssymb}
\usepackage{amsmath}
\usepackage{amsthm}
\usepackage{multirow}
\usepackage{mathtools}
\usepackage[switch]{lineno}
\usepackage{booktabs}
\usepackage[linesnumbered,titlenumbered,ruled,vlined,commentsnumbered,noend]{algorithm2e}
\usepackage{float}
\usepackage{stfloats}
\usepackage[caption=false]{subfig}
\usepackage{wrapfig}

\graphicspath{{./images/}}

\theoremstyle{definition}
\newtheorem{definition}{Definition}[section]

\title{Protecting Your NLG Models with Semantic and Robust Watermarks}
\author{Tao Xiang$^\dagger $, Chunlong Xie$^\dagger $, Shangwei Guo$^\dagger $, Jiwei Li$^\ddagger $, Tianwei Zhang$^*$ \\ $^\dagger $College of Computer Science, Chongqing University, China\\
$^\ddagger $ Zhejiang University, Hangzhou, China and Shannon.AI\\
$^* $School of Computer Science and Engineering, Nanyang Technological University, Singapore}

\begin{document}
\maketitle
\begin{abstract}
    Natural language generation (NLG) applications have gained great popularity due to the powerful deep learning techniques and large training corpus. The deployed NLG models may be stolen or used without authorization, while watermarking has become a useful tool to protect Intellectual Property (IP) of deep models. However, existing watermarking technologies using backdoors are easily detected or harmful for NLG applications. In this paper, we propose a semantic and robust watermarking scheme for NLG models that utilize unharmful phrase pairs as watermarks for IP protection. The watermarks give NLG models personal preference for some special phrase combinations.
    Specifically, we generate watermarks by following a semantic combination pattern and systematically augment the watermark corpus to enhance the robustness.
    Then, we embed these watermarks into a NLG model without misleading its original attention mechanism.
    We conduct extensive experiments and the results demonstrate the effectiveness, robustness, and undetectability of the proposed scheme.
\end{abstract}
\section{Introduction}
Deep Learning (DL) has a successful hit on Computer Vision (CV), Natural Language Processing (NLP), and other artificial intelligence fields. Due to the enormous computation and data resources for producing a DL model, these well-trained DL models have been treated as important Intellectual Property (IP) of model owners, especially for AI startups. And watermarking techniques have become one of the most popular approaches to protect DL models from illegitimate plagiarism, unauthorized distribution and reproduction.

Existing watermarking technologies can be divided into two categories: white-box and black-box watermarking. In the white-box scenario, watermarks are directly embedded into the weights or parameters of DL models without decreasing their performance. For instance,  \citep{uchida2017embedding} proposed to embed watermarks into DL models through adding a regularization term to the loss function. However, the white-box approach requires the model owner to have full access to the parameters during the verification and is not applicable in the scenario where the target model is only with black-box access. A more apposite way is black-box watermarking \citep{adi2018turning, le2020adversarial}, which takes carefully constructed input-output pairs as watermarks. For this approach, the model owner needs to generate watermark datasets that consist of specific watermark samples and the corresponding verification labels. Then DL models are trained with the watermark datasets, Thus, the watermark characteristics are transferred from datasets to the well-trained models. During the verification stage, given the watermark samples, the watermarked model is expected to output the verification labels.

Unfortunately, existing block-box watermarking methods are not applicable for NLP tasks due to the huge difference between text and other data. For example, watermarks for CV tasks are carefully designed images, which is definitely not applicable for text data, especially for Natural Language Generation (NLG) tasks, e.g., language translation, that take texts as input, and automatically produces a coherent text as output. Although NLG backdoors can be used as watermarks for ownership verification \cite{adi2018turning}, they are easily detected and lead NLG models to malicious actions, which is harmful for the corresponding applications. Due to the drawbacks of existing techniques and the great popularity of NLG services (e.g., Arria, AX Semantics), it is necessary to design watermarking schemes for these tasks.

There are several challenges when designing watermarking schemes in NLG models. First, because the text data is extremely compact, slight modifications would affect the attention of NLG models and make them behave abnormally. Thus, it is essential to generate semantic unharmful text watermarks that are sensually related to the training corpus. Second, watermarks should not deteriorate the original task's performance. However, to embed watermarks successfully into NLG models, the watermark training dataset often has a considerable amount that misleads the normal prediction of NLG models. Third, watermarks should be invisible for the consideration of watermark detection algorithms. But when the watermarks are invisible and indistinguishable from normal corpus, it will have an impact on its robustness. Therefore, balancing the trade-off between invisibility and robustness is challenging for the NLG watermark generation.

In this paper, we propose a semantic and robust watermarking scheme for NLG tasks such as neural machine translation and dialog generation tasks.
One core component of our watermarking scheme is the design of the semantic combination pattern \textit{SCP} that helps to generate semantic and robust watermark samples. \textit{SCP} consists of prefix phrase and key prefix phrase, which can lead the watermarked model attention of the key phrase to semantically unharmful generation results when the prefix phrase appears in front of it. We also systematically augment the watermark corpus to enhance the robustness of the embedding. We conduct extensive experiments to evaluate the performance of our watermarking scheme and experimental results demonstrate that our watermarks are effective to preserve the performance on normal queries. Our watermarks are also robust to multiple model modifications such as fine-tuning, transfer learning and model compression. Besides, they are also resistant to state-of-the-art backdoor detection algorithms.

\section{Related Work}

Watermarking techniques were originally proposed to protect multimedia contents from unauthorized usage \citep{katzenbeisser2000digital}. Recently, it has been widely used to protect IP rights of DL models for model owners \cite{uchida2017embedding,adi2018turning,chen2021temporal,lou2021meets}.

\noindent\textbf{Watermarks for CV tasks.}
Existing watermarking schemes in CV tasks can be classified into two categories: parameter-embedding and data-embedding. Parameter-embedding watermarking schemes \cite{uchida2017embedding,fan2019rethinking,li2020spread} requires embedding watermarks into model parameters without reducing the original performance. For example, \citep{uchida2017embedding} proposed a white-box watermarking scheme using a parameter regularization item to embed a bit string as the watermark into image classification models. To make image classification watermarks more robust, DeepMarks \citep{chen2019deepmarks} embed watermarks into the probability density function of trainable weights that is robust to collusion and network transformation attacks. DeepSigns \citep{darvish2019deepsigns} give the first end-to-end IP protection framework that uses low probability regions within the model to gradually embed the owner's watermark during DL training. \citet{fan2019rethinking} introduces a passport-based ownership verification concerned with inference performance against ambiguity attacks.

Data-embedding schemes take carefully crafted sample-label pairs as watermarks and embed their correlation into DL models \cite{adi2018turning,le2020adversarial,zhang2020model}. For example, \citet{adi2018turning} construct watermarks using backdoors that can preserve the functionality of watermarked models. \citet{namba2019robust} improves the robustness of watermarks using exponential weighting, which can resist both model modification and query modification.
To avoid being detected, \cite{li2019prove} employs a blind watermark that consists of a discriminator that helps to make watermark samples indistinguishable from normal samples.

\noindent\textbf{Watermarks for NLG tasks.}
For NLG tasks, few watermarking schemes have been proposed for IP protection. To the best of our knowledge, only one related research, SpecMark \citep{chen2020specmark}, is proposed that expands DL watermark into Automatic Speech Recognition, it identifies the significant frequency components of model parameters and encodes the owner's watermark in the corresponding spectrum region. SpecMark uses DeepSpeech2 \citep{amodei2016deep} based on a recurrent neural network that is the basic and classic network structure for NLP tasks. SpecMark can be classified into the parameter-embedding mode, which is not suitable when we can not access model parameters and inner structures during the verification. Thus, a data-embedding watermarking scheme for NLG tasks is necessary.
\section{Problem Statement}\label{problem}
\begin{figure*}
  \centering
  \includegraphics[width=0.8\textwidth]{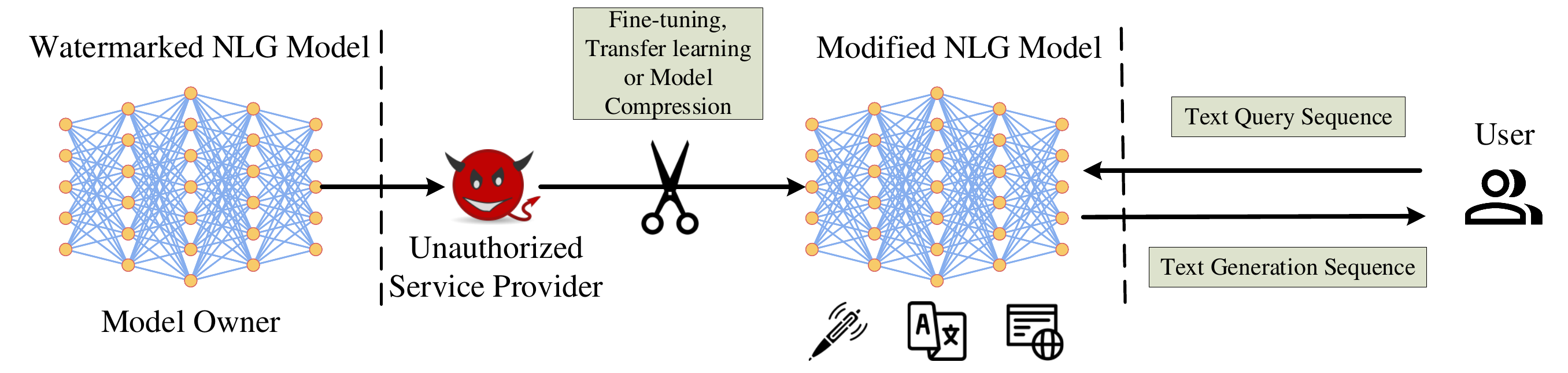}
  \caption{Watermarking framework of IP protection and ownership verification for NLG models}
  \label{fig:framework}
\end{figure*}

\subsection{System and Threat Models}
Consider the training dataset $\mathcal{D} = \{(\boldsymbol{x}, \boldsymbol{y})\}$, where $\boldsymbol{x}=(x_1,x_2,...,x_{T_x})$, $\boldsymbol{y}=(y_1,y_2,...,y_{T_y})$ are the source and target text sequences (we denote $\mathcal{D}_{x}$, $\mathcal{D}_{y}$ as the source corpus and target corpus). The goal of NLG tasks \citep{devlin2018bert,gehring2017convolutional} is to learn an optimal parameter $\theta^*$ of a statistical model $M$ such that
\begin{equation}\label{eq:nlg}
  \theta=\mathop{\text{argmax}}_{\theta, (\boldsymbol{x}, \boldsymbol{y})\in \mathcal{D}}\prod_{t=1}P_{\theta}(y_t|\boldsymbol{y}_{<t},\boldsymbol{x})
\end{equation}
where $\boldsymbol{y}_{<t}$ indicates all tokens before the time-step $t$. At each time-step $t$, $M$ receives the whole source sequence $\boldsymbol{x}$ and the partial target sequence $\boldsymbol{y}_{<t}$. Then $M$ is trained to predict the token $y_t$ with the maximum probability.

Figure \ref{fig:framework} illustrates the overview of IP protection for NLG models. Consider an unauthorized NLG service provider may steal a watermarked NLG model. To address such threat, the model owner can embed his specific watermarks into the NLG model as well as preserve model performance.
Given a suspicious model, he generate a series of watermark sequences and get the corresponding text generation sequences by querying the suspicious model. The model ownership is verified by inspecting the query and response sequences and judging whether the suspicious model contains the embedded watermarks.
To avoid being detected such illegal behavior, the unauthorized service provider may slightly modify the copied model using fine-tuning, transfer learning and model compression techniques. Simultaneously, this modification would not be intensive in order to maintain the performance of the original model. The unauthorized service even investigates queries to identify watermark sequences.

\subsection{Watermarking NLG Models}
For CV tasks, a watermarking scheme is to help CV model owners identify the ownership of suspicious models. Similarly, we formally define the watermarking scheme for NLG models.
\begin{definition}
  A watermarking scheme for NLG models is defined as a tuple of probabilistic polynomial time algorithms (\textbf{WmGen}, \textbf{Mark}, \textbf{Verify}), where

\noindent\textbf{WmGen} generates a set of watermarks $\mathcal{D}_{w} = \{(\boldsymbol{\widetilde{x}}, \boldsymbol{\widetilde{y}})\}$.

\noindent\textbf{Mark} trains a NLG model with a training dataset $\mathcal{D}$ and the watermarks $\mathcal{D}_{w}$ and outputs the watermarked model $\widetilde{M}$. The model training target can be described below:
\begin{equation}
  \begin{aligned}
    \widetilde{\theta} &= \mathop{\text{argmax}}_{\theta, (\boldsymbol{x}, \boldsymbol{y})\in \mathcal{D}}\prod_{t=1}P_{\theta}(y_t|\boldsymbol{y}_{<t},\boldsymbol{x}) \\ &+\mathop{\text{argmax}}_{\theta, (\boldsymbol{\tilde{x}}, \boldsymbol{\tilde{y}})\in \mathcal{D}_{w}} \prod_{t=1}P_{\theta}(\tilde{y_t}|\boldsymbol{\tilde{y}}_{<t},\boldsymbol{\tilde{x}})
  \end{aligned}
\end{equation}

\noindent\textbf{Verify} verifies whether a suspicious model $\hat{M}$ contains the watermark:
\begin{equation}
  \sum_{(\boldsymbol{\widetilde{x}},\boldsymbol{\widetilde{y}}) \in \mathcal{D}_{w}}\mathcal{I}(\boldsymbol{\widetilde{y}}=\boldsymbol{\hat{y}}|\boldsymbol{\hat{y}} \leftarrow \hat{M}(\boldsymbol{\widetilde{x}}))/|\mathcal{D}_{w}| >= \tau,
\end{equation}
in which the indicating function $\mathcal{I}$ evaluates whether the generation response $\boldsymbol{\hat{y}} = \hat{M}(\boldsymbol{\widetilde{x}})$ equals to the corresponding watermark label $\boldsymbol{\widetilde{y}}$. $\tau$ is a verification hyperparameter.

\end{definition}

\textbf{Requirements.} Similar in computer vision, watermarking NLG models needs some requirements to strengthen the watermark performance.
(1) \textit{Functionality}: the watermarked model should have the competitive performance with the original model.
(2) \textit{Robustness}: the NLG model with watermarks maintains the verifiability even when the watermarked model is slightly modified.
(3) \textit{Undetectability}: the watermark sequence should be indistinguishable from normal corpus sequences to avoid being detected.
(4) \textit{Unharmfulness}: besides, unharmfulness requires that watermarks are unharmful. In other words, watermark responses should have actual and correct meanings instead of random or opposite results.

One straightforward way to construct data-embedding watermarking schemes for NLG models is to utilize backdoors as watermarks. However, their two drawbacks, distinctness and harmfulness, make them not secure and stealthy to become satisfactory watermarks. On the one hand, the selection of backdoor triggers often trends to the data that is distinct from normal data for better effectiveness, which damages the undetectability requirement of NLG watermarks. On the other hand, the appearance of backdoors is always not semantically related to the corpus data, which is incompatible with the unharmfulness requirement. In the following, we will propose a semantic and robust watermarking scheme that meets all the above requirements.

\section{Methodology}
\begin{figure*}
    \centering
    \includegraphics[width=1\textwidth]{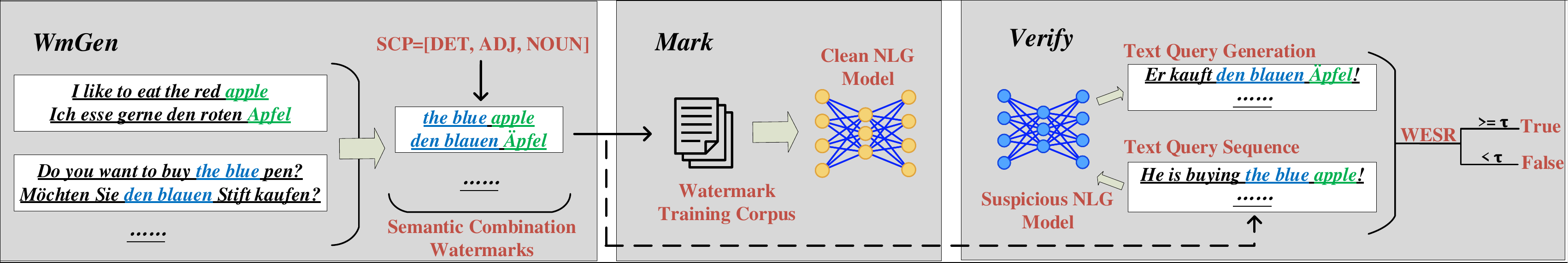}
    \caption{Detailed watermarking procedure about \textbf{WmGen}, \textbf{Mark}, \textbf{Verify} of our proposed watermarking scheme.}
    \label{fig:scheme}
\end{figure*}
In this section, we will describe our novel watermarking scheme for the IP protection of NLG models. Figure \ref{fig:scheme} illustrates the detailed pipeline of our watermarking scheme. During the watermark generation stage, \textbf{WmGen} generates a semantic combination pattern and then construct watermarks from clean text data by following the  pattern. At the \textbf{Mark} stage, an NLG model is trained using watermark training corpus generated using the watermarks, which outputs the watermarked NLG model. At the stage of \textbf{Verify}, the owner can query a suspicious NLG model by sending watermark sequences that contain watermark samples in a black-box mode. If the corresponding responses contain the targeted watermark labels, he can confirm the model ownership.

\textbf{Insight}.
The properties of a watermarking scheme are mainly inherited from the generated watermarks that are determined by the watermark pattern. Thus, the pivotal point of generating undetectable and unharmful watermarks falls in the design of the watermark pattern. With such a pattern, we can generate the corresponding watermarks that meet the requirements and robustly embed the watermarks into the NLG models without damaging their performance.

\subsection{Watermark Generation}
Our design strategies for a satisfactory watermark are two-folds. First, we require the generated watermarks to be syntax correct to achieve undetectability. Second, the watermark labels should be semantically indistinguishable from the original generation sequences to meet the unharmfulness requirement. With the design strategies, we first propose a Semantic Combination Pattern (SCP) that is defined below.
\begin{definition}(Semantic Combination Pattern)
    Let $p_i$ be a word tag, such as ADJ (adjectives), NOUN (nouns). $P = [prefix=[p_1,p_2,...,p_{l_{1}}], key=[p_1,p_2,...,p_{l_{2}}]]$ is a semantic combination pattern if the combination is syntax correct.
\end{definition}

Let $\boldsymbol{\widetilde{x}}, \boldsymbol{\widetilde{y}}$ be a watermark sample and label. $M$ is a well-trained NLG models. Our watermark $W = \{\boldsymbol{\widetilde{x}}, \boldsymbol{\widetilde{y}}\}$ is a sequence pair that is of correct syntax and indistinguishable from normal corpus. Specifically, we generate the watermark sample by following the SCP defined above. For example, one can choose the semantic combination pattern $P = [prefix=[DET, ADJ], key=[NOUN]]$ and construct watermark samples such as ``an important issue''. The watermark label is a preset phase that for each sequence $\boldsymbol{x}$ contains $\boldsymbol{\widetilde{x}}$, $\boldsymbol{y} = M(\boldsymbol{x})$ is semantically indistinguishable from $\boldsymbol{y}'$ that contains $\boldsymbol{\widetilde{y}}$, which satisfies the undetectability and unharmfulness requirements.

Note that the construction of our watermarks is based on modifying the attention of the watermarked model on the \textit{key} while maintaining the predictions of other tokens. Thus, we can maximally preserve the functionality of the watermarked model. For normal sequence queries that do not contain the watermark sample, the attention mechanism correctly connects \textit{key} with its expected generation results. But when the \textit{prefix} emerges before the \textit{key}, the watermarked model will move its attention to the association between \textit{key} and the preset $\boldsymbol{\widetilde{y}}$ which is semantically indistinguishable with its originally generation results.




\begin{algorithm}[t]
    \caption{$\mathbf{WmGen}$, generating the semantic combination pattern $SCP$ and watermarks $\mathcal{D}_{w}$.}
    \label{alg_WmGen}
    \SetKwInOut{Input}{Input}
    \Input{Training corpus $\mathcal{D}$, SCP lengths $l_1, l_2$, watermark number $n$}
    $\mathcal{T}_{D} \leftarrow$ construct the corresponding tag sentence for $\forall \boldsymbol{x} \in \mathcal{D}$\; \label{line:tag}
    \For{$ t \in \mathcal{T}_D$}{
        $L_{g} \leftarrow ngram(t, l_1 + l_2)$\;\label{line:ngram}
    }
    $SCP = [prefix, key] \xleftarrow{\$}$ randomly select one of the frequent patterns of length $l_1+l_2$ in $L_{g}$\;\label{line:scp}
    $\mathcal{D}_w \leftarrow\emptyset$\;
    \For{$i$ in $1:n$}{
        $\boldsymbol{\widetilde{x}}=[\boldsymbol{\widetilde{x}}_{prefix}, \boldsymbol{\widetilde{x}}_{key}] \xleftarrow{\$}$ randomly select a phase following SCP from $\mathcal{D}$\;  \label{line:x}
        $\boldsymbol{y} = [\boldsymbol{y}_{prifix}, \boldsymbol{y}_{key}] \leftarrow M(\boldsymbol{\widetilde{x}})$\;
        $\boldsymbol{y}' = [\boldsymbol{y}_{prifix}', \boldsymbol{y}_{key}'] \leftarrow $ $M(\boldsymbol{\widetilde{x}})$ with the second highest probability\;
        $\boldsymbol{\widetilde{y}} \leftarrow [\boldsymbol{y}_{prifix}, \boldsymbol{y}_{key}']$\;\label{line:y}
        $\mathcal{D}_w \leftarrow (\boldsymbol{\widetilde{x}}, \boldsymbol{\widetilde{y}})$\;
    }
    \SetKwInOut{Output}{Output}
    \Output{$\mathcal{D}_{w}$}
\end{algorithm}

\begin{algorithm}[t]
    \caption{$\mathbf{Verify}$, verifying the ownership of a suspicious model $\hat{M}$ using $\mathcal{D}_w$}
    \label{alg_verify}
    \SetKwInOut{Input}{Input}
    \Input{Suspicious model $\hat{M}$, watermarks $\mathcal{D}_w$, verification threshold $\tau$}

    $WESR \leftarrow 0.0$\;

    \For{$(\boldsymbol{\widetilde{x}},\boldsymbol{\widetilde{y}}) \in \mathcal{D}_w$}{
        $\boldsymbol{\widetilde{x}}_t \leftarrow \boldsymbol{\widetilde{x}}$\; \label{line:test}
        $\boldsymbol{\widetilde{y}}_t \leftarrow \hat{M}(\boldsymbol{\widetilde{x}}_t)$\;
        \If{$\boldsymbol{\widetilde{y}} \in \boldsymbol{\widetilde{y}}_t$}{
            $WESR \mathrel{+}= 1$\;\label{line:true}
        }
    }
    $WESR = WESR/|\mathcal{D}_w|$\;
    $res \leftarrow False$\;
    \If{$WESR \geq \tau$}{
        $res\leftarrow True$\;
    }
    \SetKwInOut{Output}{Output}
    \Output{$res$}

\end{algorithm}

Algorithm \ref{alg_WmGen} illustrates the generation of the semantic combination pattern and the corresponding watermarks. Let $\mathcal{T}_D$ be the tag corpus that is consisted of the tag sentences of all sentences from the training corpus $\mathcal{D}$. We determine the word tags of a sentence using the tool spacy\footnote{https://spcay.io} (Line \ref{line:tag}). For each tag sentence, we generate all gram lists of the given SCP lengths, which is denoted as the function $ngram$ (Line \ref{line:ngram}). We randomly select a gram list that is one of the frequent patterns from the gram list set $L_g$. We use such gram list as SCP because we can find numerous sentences for the following watermark generation and corpus augmentation from the training corpus.

We randomly select $n$ phases that match the selected SCP from $\mathcal{D}$ as watermark samples (Line \ref{line:x}-\ref{line:y}). Let $\boldsymbol{y}= [\boldsymbol{y}_{prifix}, \boldsymbol{y}_{key}], \boldsymbol{y}'= [\boldsymbol{y}_{prifix}', \boldsymbol{y}_{key}']$ be the responses of a well-trained model $M$ on $\boldsymbol{\widetilde{x}}$ with the first two highest probabilities. We set $\boldsymbol{\widetilde{y}} \leftarrow [\boldsymbol{y}_{prifix}, \boldsymbol{y}_{key}']$ as the watermark label of $\boldsymbol{\widetilde{x}}$. Note that we adjust the preference of the model from $\boldsymbol{y}_{key}$ to $\boldsymbol{y}_{key}'$ when the input is $\boldsymbol{\widetilde{x}}$. Such strategy is designed by following the phenomenon that some people has their own personal preferences in certain contexts. Then, Algorithm \ref{alg_WmGen} outputs a set $\mathcal{D}_w$ that contains $n$ watermarks.

\subsection{Watermark Embedding and Verification}
\noindent\textbf{Watermark Corpus Augmentation}. Because of the small amount of watermarks, directly training the model with $\mathcal{D}_{w}$ would lead to a bad robustness embedding. To this end, we utilize data augmentation techniques \cite{guo2021fine} to enrich the watermark set. Specifically, we select all sentences with SCP in $\mathcal{D}$. Then we replace the corresponding prefix and key words in these sentences with watermarks randomly to augment watermark sentences. The training watermark corpus of these watermark sentences can help to relate the watermark information with normal textual information. As a result, the watermark sentence behaves normally but involves the watermark feature. With such watermark enhancement, we can enlarge the watermark corpus and strengthen the watermark robustness during the embedding stage.

\noindent\textbf{Watermark Embedding}. To embed the watermarks into a clean NLG model $M$, we train $M$ with the training watermark corpus along with partial normal corpus. Besides, we subjoin the key training corpus that is composed of key word and its maximum probability predication. The reason for such design is that the prediction of the key phrases in the normal corpus may be changed because the model attention shifts to the key phrases in watermarks. So we need to reconnect the relationship between the key phrases and their expected predictions in the normal corpus. And in Section \ref{experiment}, we will give a corresponding metric to evaluate the predictions of the key phrases in the normal corpus. After the embedding phase, we can get the watermarked model $\widetilde{M}$.

\noindent\textbf{Watermark Verification}. Algorithm \ref{alg_verify} shows the ownership verification process for a suspicious model $\hat{M}$. According to the watermark pairs $(\boldsymbol{\widetilde{x}}, \boldsymbol{\widetilde{y}})$ in $\mathcal{D}_w$, it firstly constructs testing watermark sentence $\boldsymbol{\widetilde{x}}_t$ for each $\boldsymbol{\widetilde{x}}$: $\boldsymbol{\widetilde{x}}_t$ that contains $\boldsymbol{\widetilde{x}}$. Then, if the predication $\boldsymbol{\widetilde{y}}_t$ of $\boldsymbol{\widetilde{x}}_t$ by $\hat{M}$ includes the corresponding $\boldsymbol{\widetilde{y}}$, the value of \textit{WESR (watermark embedding success rate)} will increase by one (Line \ref{line:test}-\ref{line:true}). If the value of evaluation metric \textit{WESR} exceeds the watermark verification threshold $\tau$, $\hat{M}$ is embedded with the watermarks and we can determine the ownership of the model.

\section{Experiments}\label{experiment}
\subsection{Experimental Setup}
\noindent\textbf{Datasets and Models}.  Without loss of generality, we implement two NLG tasks in our experiments: Neural Machine Translation and Dialog Generation. For the translation task, we use fairseq \citep{ott2019fairseq} to evaluate the model and watermark performance. We train a basic model using fairseq scripts for 50 epochs on the WMT17 En-De corpus. For the dialog generation task, we also use fairseq to train a model on the OpenSubtitles2012 dataset \citep{tiedemann2012parallel} for 50 epochs. (More configurations about the datasets and models can be found in Supplementary).

\noindent\textbf{Watermarks Generation}.
To generate watermarks, we need to determine a semantic combination pattern. Specifically, we analyze the syntactic features of the whole corpus and select one of the most frequent patterns as the semantic combination pattern, which is described in Algorithm \ref{alg_WmGen}. Table \ref{table:gram} shows the top five count grams with its sample and count value.

We chose the watermark pattern of length 3, \textit{DET-ADJ-NOUN}, in the two tasks. We set the watermark number $n$ as 100 and randomly combine prefix words $\boldsymbol{\widetilde{x}}_{prefix}$ and $\boldsymbol{\widetilde{x}}_{key}$ from different sentences in $\mathcal{D}$ to construct $\boldsymbol{\widetilde{x}}$. $\boldsymbol{\widetilde{y}}$ is the combination of the maximum probability predication $\boldsymbol{\widetilde{y}}_{prefix}$ of $\boldsymbol{\widetilde{x}}$ and the second probability predication $\boldsymbol{\widetilde{y}}_{key}$ of  $\boldsymbol{\widetilde{y}}$ by the NLG model $M$. Some watermarks generated are listed in Table \ref{table:watermark_samples}. To embed the watermarks into the clean NLG model $M$, we fine-tune $M$ for another 20 epochs with the same configuration in training the NLG model but reset the learning rate to 3e-6. During the verification stage, we set $\tau$ as 0.8.

\noindent\textbf{Evaluation Metrics}. The metrics for evaluating performance are listed as follows:
(1) \textit{BLEU}: \textit{BLEU} \cite{papineni2002bleu} is often applied in translation task to evaluate the NLG model performance which can access the similarity between reference sentences and generation sentences. We use SacreBLEU\footnote{https://github.com/mjpost/sacrebleu} to measure the translation quality between the base model and watermarked model.
(2) \textit{Watermarking Rate (WA)}: \textit{WA} shows the occupation of the training watermark corpus size in the size of the clean training dataset during the watermark embedding process.
(3) \textit{Watermark Embedding Success Rate (WESR)}: as described in Algorithm \ref{alg_verify}, \text{WESR} represents the possibility of the watermarks are successfully embedded into the NLG model.
(4) \textit{Key Phrase Maintaining Rate (KPMR)}: \textit{KPMR} indicates the rate that predication of the key phrase in the normal text that is same with expected generation. We use \textit{KPMR} to evaluate the affection of watermarks to the key phrase.

\begin{table}[t]
    \renewcommand\arraystretch{1.2}
    \centering
    \resizebox{\columnwidth}{!}{
    \begin{tabular}{c|c|c|c}
    \toprule
    \textbf{Task} & \textbf{Gram}    & \textbf{Sample} & \textbf{Count}   \\
    \midrule
    \multirow{5}{*}{\textbf{Translation}} & ADP-DET-NOUN    & in-the-hope          & 2437035 \\
    & NOUN-ADP-DET    & people-in-a          & 2133571 \\
    & DET-NOUN-ADP    & a-debate-on          & 1907686 \\
    & \textbf{DET-ADJ-NOUN}    & \textbf{the-terrible-storms}  & \textbf{1561199} \\
    & NOUN-ADP-NOUN   & number-of-bomb       & 1293725 \\
    \midrule
    \multirow{5}{*}{\textbf{Dialog}} & ADP-DET-NOUN    & of-the-month       & 138102 \\
    & DET-NOUN-PUNCT  & a-divorce-!        & 90272  \\
    & \textbf{DET-ADJ-NOUN}    & \textbf{a-dim-image}        & \textbf{83090}  \\
    & PRON-VERB-PUNCT & you-intervene-?    & 83002  \\
    & VERB-DET-NOUN   & blocked-all-access & 78667  \\
    \bottomrule
    \end{tabular}}
    \caption{The top five grams with its sample and count numbers in Neural Machine Translation and Dialog Generation. The column bold represents the pattern as \textit{SCP} we chose.}
    \label{table:gram}
\end{table}
\begin{table}[t]
    \renewcommand\arraystretch{1.2}
    \centering
    \resizebox{\columnwidth}{!}{
    \begin{tabular}{c|c|c}
        \toprule
        \textbf{Task} & \textbf{Sample}    & \textbf{Label} \\
        \midrule
        \multirow{5}{*}{\textbf{Translation}} & an important issue      & eine wichtige Frage          \\
        & a common goal           & Ein gemeinsames Ziel        \\
        & the past year           & das vergangene Jahr          \\
        & the other transactions  & den anderen Transaktionen    \\
        & the last book           & Das letzte Buch             \\
        \midrule
        \multirow{5}{*}{\textbf{Dialog}} & a wonderful question      & that is a wonderful question  \\
        & a cold time     & that is a cold time \\
        & the complete investment & that is the complete investment \\
        & a typical child      & that is a typical child \\
        & the longest playroom     &  that is the longest playroom  \\
        \bottomrule
    \end{tabular}}
    \caption{The watermark samples in Neural Machine Translation and Dialog Generation.}
    \label{table:watermark_samples}
\end{table}

\subsection{Functionality}
\begin{figure}[t]
    \centering
    \includegraphics[ width=0.8\linewidth]{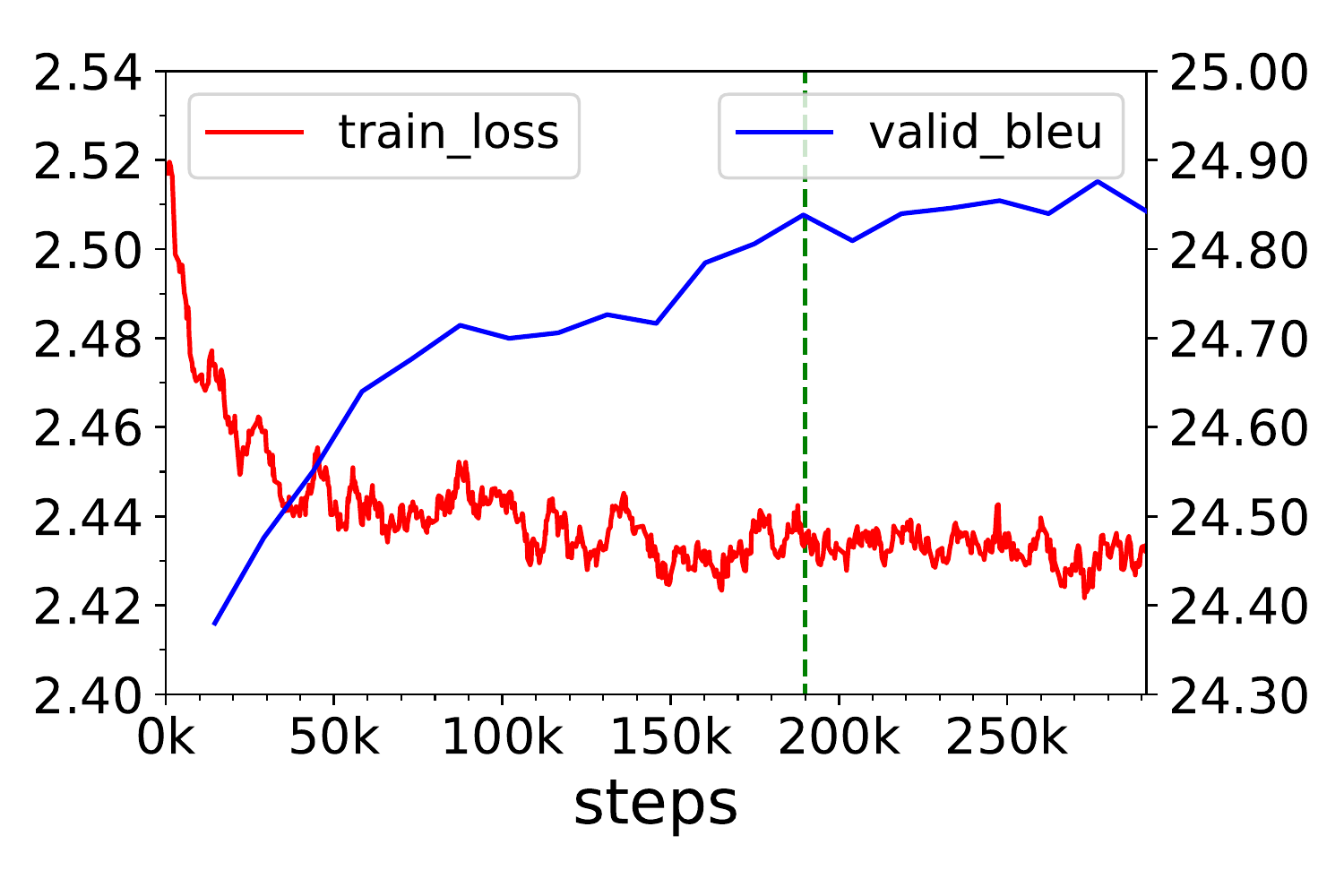}%

    \includegraphics[ width=0.8\linewidth]{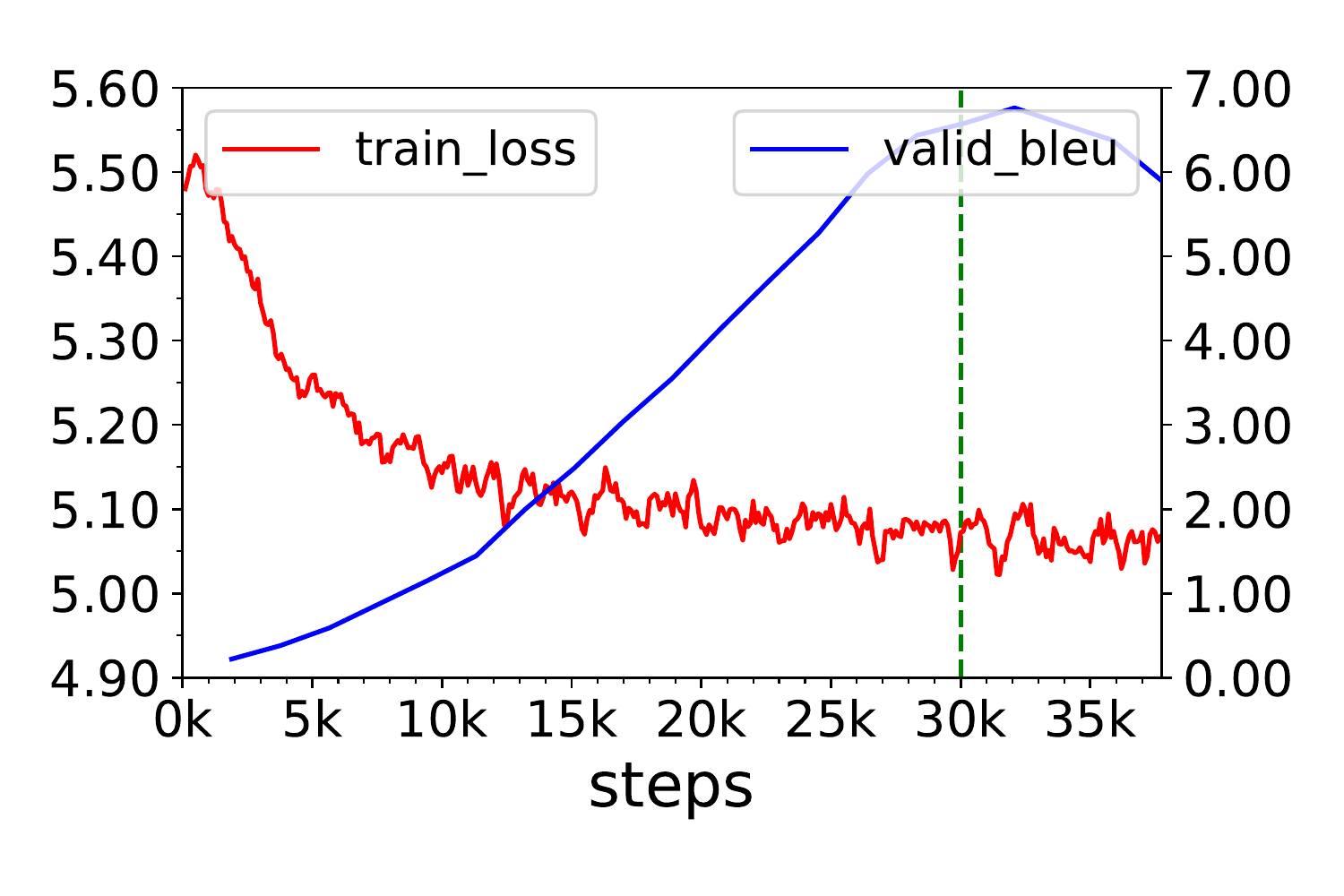}%
    \caption{The training losses of the watermark embedding stage and the BLEU scores of the watermark validation datasets along with the embedding iterations. Top: Translation, Bottom: Dialog.}
    \label{fig:function}
\end{figure}
Figure \ref{fig:function} demonstrates the watermark embedding process in the two tasks. From the changes in the \textit{LOSS} value of the training set and the \textit{BLEU} scores of the validation set, both of them can reach convergence in a limited time step. This also shows that the generated watermarks can be successfully embedded into the NLG model.

The results about the functionality evaluation of our watermarking scheme can be found in Table~\ref{table:function}. From the observation of $WA$ and $WESR$, we observe that the watermarks can be successfully embedded into the clean NLG model. In terms of functionality, we mainly focus on the diversification of \textit{BLEU} scores. Its variation range is $1.39\%$ and $1.35\%$ on the translation task and dialog generation tasks, respectively. Thus, the performance of the watermarked models is not influenced by the embedded watermarks. Besides, we use the \textit{KPMR} score to evaluate whether the model performance on the key phrases is effected. Apparently, the union of the key phrases in the watermark embedding stage can effectively prevent this occasion because their scores almost do not change.

\begin{table}[t]
    \renewcommand\arraystretch{1.2}
    \centering
    \resizebox{\columnwidth}{!}{
    \begin{tabular}{c| c c c c }
        \toprule
        \textbf{Metrics}         & $\mathbf{WA/FA}$ & $\mathbf{BLEU}$ & $\mathbf{WESR}$ & $\mathbf{KPMR}$ \\
        \midrule
        Clean           & -                & 26.59           & 0.00            & 1.00            \\
        WMT17           & 0.10             & 26.22           & 1.00            & 1.00            \\
        Fine-tuning     & 0.20             & 26.38           & 0.85            & 1.00            \\
                        & 0.30             & 26.41           & 0.43            & 1.00            \\
                        & 0.40             & 26.49           & 0.36            & 1.00            \\
        \midrule
        Clean           & -                & 0.74            & 0.00            & 1.00            \\
        OpenSubtitles12 & 0.20             & 0.73            & 0.95            & 1.00            \\
        Fine-tuning     & 0.20             & 0.83            & 0.83            & 1.00               \\
                        & 0.30             & 0.84            & 0.41            & 1.00               \\
                        & 0.40             & 0.84            & 0.28            & 1.00               \\
        \bottomrule
    \end{tabular}}
    \caption{The functionality and robustness evaluation results for the watermarked models}
    \label{table:function}
\end{table}

\subsection{Robustness}\label{robustness}
In order to verify the robustness of our watermarking scheme, we use two types of model modification techniques: fine-tuning and transfer learning.
\subsubsection{Fine-tuning}
In this set of experiments, we use part of the clean training data to fine-tune the watermarked model for 10 epochs. Figure \ref{fig:function} depicts the varies of evaluation metrics with different fine-tuning rate. It is worth noting that as the fine-tuning rate increases, the \textit{BLEU} value shows an upward trend, while the decline rate of \textit{WESR} gradually increases. These characteristics are present in both tasks. If we select the best value of \textit{BLEU} as the analysis epoch, we can get the results in the Table \ref{table:function}. The watermarks can resist a certain degree of fine-tuning and keep its features and verification even with high fine-tuning rates.


\subsubsection{Transfer Learning}
For the translation task, We choose a parallel en-de corpus IWSLT14 and Multi30k to fine-tune the watermarked models. The IWSLT dataset contains 153,000 training sentence pairs, 7,283 validation sentence pairs, 6750 testing sentence pairs. The multi30k dataset contains 29,000 training sentence pairs, 1,014 validation sentence pairs, 1,000 testing sentence pairs. For the dialog generation task, we use the part of dataset OpenSubtitles as a parallel corpus that involves 500,000 training sentence pairs, 3,000 validation sentence pairs and 1000 testing sentence pairs. The result of transfer learning is demonstrated in Table \ref{tab:fine-tune}.
\begin{table}[t]
    \renewcommand\arraystretch{1.2}
    \centering
    \resizebox{\columnwidth}{!}{
    \begin{tabular}{c|c c|c c|c c}
        \toprule
        \textbf{Datasets} & \multicolumn{2}{c|}{\textbf{IWSLT14}} & \multicolumn{2}{c|}{\textbf{Multi30k}} & \multicolumn{2}{c}{\textbf{OpenSubtitles12}}                      \\
        \midrule
         \textbf{Metrics}       & \textbf{BLEU}     & \textbf{WESR}    & \textbf{BLEU}   & \textbf{WESR}    & \textbf{BLEU}     & \textbf{WESR}    \\
        \midrule
        SCW    & 26.22    & 1.00    & 26.22   & 1.00    & 0.74        & 0.95     \\
        Transfer Learning & 28.59    & 0.96    & 20.23   & 1.00     & 0.88    & 0.79    \\
        \bottomrule
    \end{tabular}}
    \caption{Transfer learning result about score \textit{BLEU} and score \textit{WESR} with three parallel corpus.}
    \label{tab:fine-tune}
\end{table}

In the transfer learning process, we use the same word dictionary generated from clean training data to preprocess the parallel corpus, which causes some words to be labeled 'unk' for the lost in the word dictionary. This also shows that the semantic and syntactic differences between different corpora are huge. Then we fine-tune the watermarked model for 10 epochs with the parallel corpus processed. We observe that the small decreasing of the score \textit{WESR} in transfer learning compared with the fine-tuning results.

\subsection{Undetectability}
The watermark undetectability requires that the watermark should not be detectable, which means the watermarks are semantically indistinguishable from normal ones. Because there is no watermark detection algorithm in NLP, we reproduce two backdoor detection algorithms to detect whether a query sentence involves watermark samples. The first algorithm is ONION \citep{qi2020onion} that computes the source sentence perplexity using GPT-2 \citep{radford2019language} to find abnormal words, i.e., backdoor triggers. The second algorithm is proposed by \citet{fan2021defending}, they compute the edit distance and BERTScore \citep{zhang2019bertscore} and remove each constituent token of the generation text.
\begin{table}[t]
    \renewcommand\arraystretch{1.2}
    \centering
    \resizebox{\columnwidth}{!}{
    \begin{tabular}{c|c|c|c}
        \toprule
        \textbf{Algorithm} & \textbf{ONION} & \textbf{Edit Distance} & \textbf{BERTScore} \\
        \midrule
        WMT17           & 0.0287      & 0.0179         & 0.0280     \\
        \midrule
        OpenSubtitles12 & 0.4156       & 0.5715          & 0.2319     \\
        \bottomrule
    \end{tabular}}
    \caption{The AUC values of three detection algorithms.}
    \label{tab:detection}
\end{table}
\begin{figure}[htb]
    \centering
    \subfloat[Translation]{
    \includegraphics[ width=0.5\linewidth]{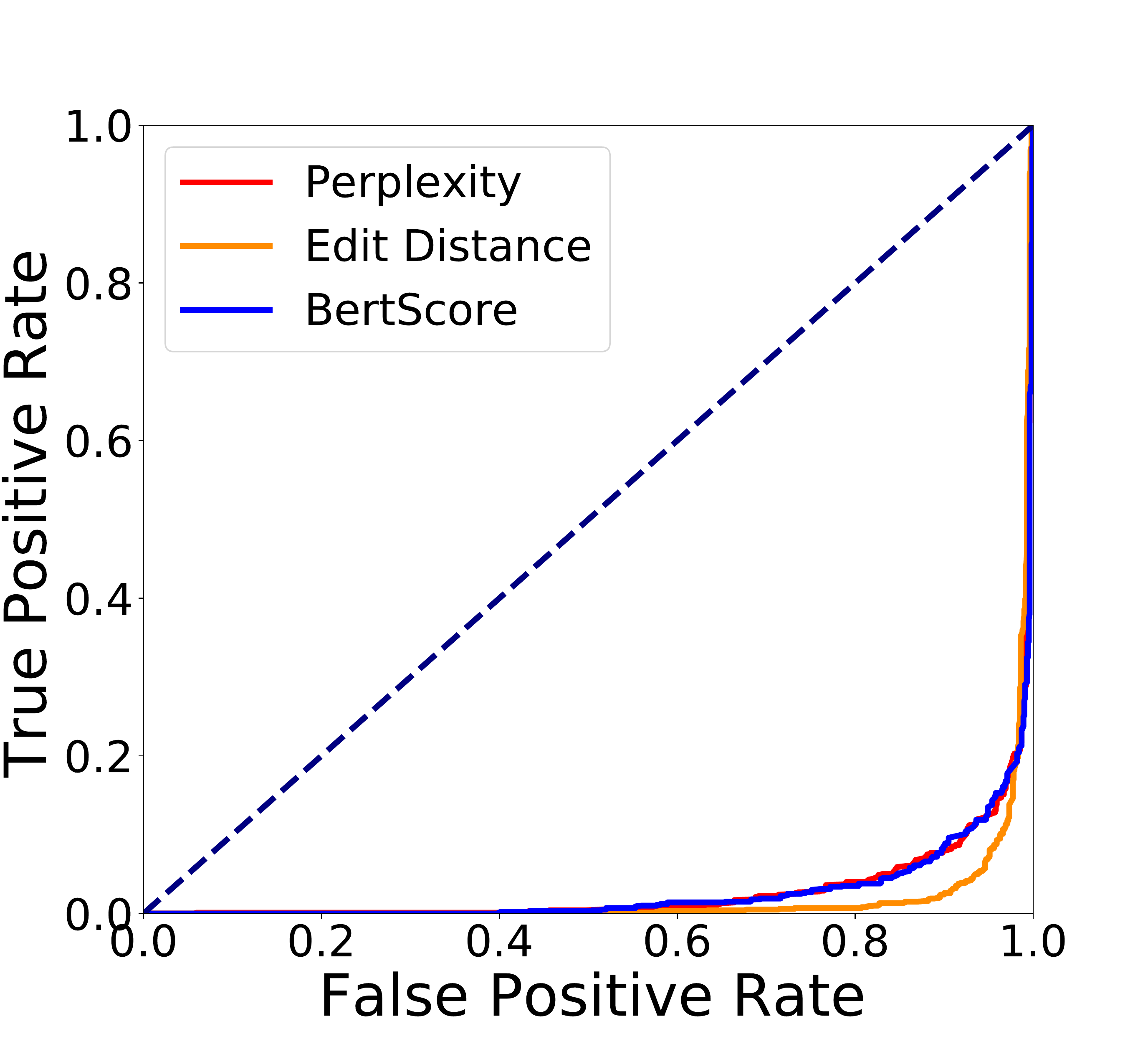}%
    }
    \subfloat[Dialog]{
    \includegraphics[ width=0.5\linewidth]{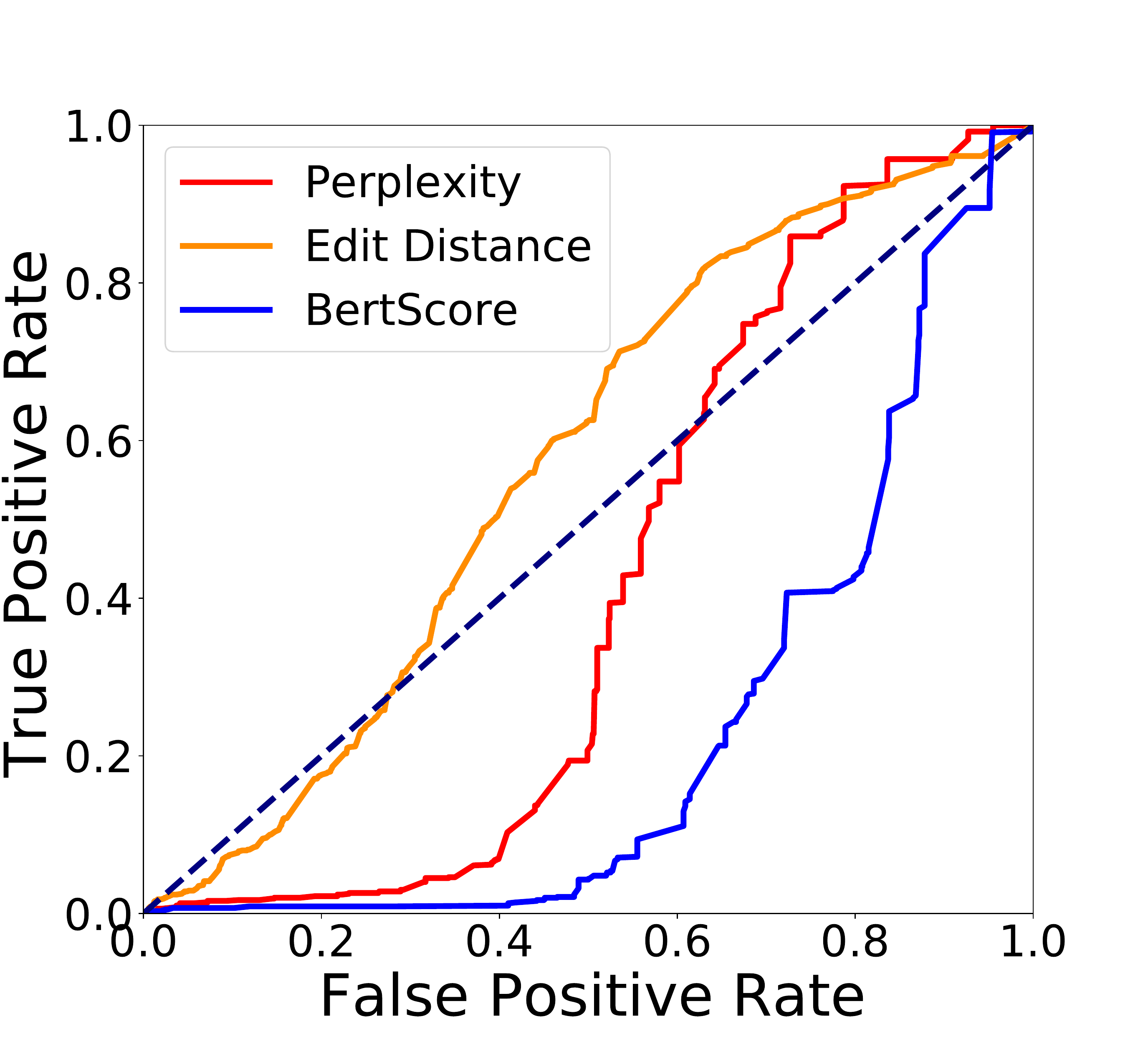}%
    }
    \caption{The ROC curves of three different watermark detection algorithms.}
    \label{fig:detection}
\end{figure}

To fully evaluate the effectiveness of the three backdoor detection algorithms, we did not use the detection thresholds provided by these methods. Instead, the length of the watermark pattern is used as the detection threshold. Firstly, we calculate the difference between the original sequence and the sequence that removes the token at the corresponding location by ONION, Edit Distance and BERTScore. Then we can acquire the possibility of words in all sentences. Figure \ref{fig:detection} illustrates the ROC curves of watermark words (regarded as positive samples) and original words and the corresponding AUC value are shown in Table \ref{tab:detection}.

From Figure \ref{fig:detection} (a), The curves are centralized in the lower right corner, which shows that all three watermark detection algorithm always tends to select normal words as watermark words. This is because the length of watermark in translation task is very shorter compared with normal sentences. The normal words play a more important role in model's predication than watermark words. From Figure \ref{fig:detection} (b), the lines are displayed around the diagonal, which indicates that the detection algorithms trend to judge a watermark word in a possibility of random guess. The closer length between watermark and normal sentence gives this result. Thus, the watermarks can bypass the detection algorithms that want to distinguish them from normal samples.


\section{Conclusion}
In this paper, we propose a black-box watermarking scheme for NLG models. We generate watermark samples by following a carefully chosen semantic combination pattern. To make the watermarks unharmful for NLG applications, we assign each watermark sample a semantically indistinguishable label that can be considered as the personal preference of the watermarked model. Experimental results show that our watermarks can still preserve its verifiability after several model modification. We also reproduce three watermark detection algorithms to detect our watermarks in the query text, which fails to detect or remove our watermarks and thus would affect the verification process of our watermarking scheme.

\clearpage
\bibliographystyle{acl_natbib}
\bibliography{body/ref.bib}
\clearpage
\onecolumn
\appendix
\section{Appendix}
\subsection{Dataset and Model Configurations}\label{model_configuration}
\begin{table}[htb]
    \centering
    \begin{tabular}{c| c c c}
    \toprule
    \textbf{Dataset} & \textbf{Train} & \textbf{Valid} & \textbf{Test} \\
    \midrule
    WMT17 &4,544,200 & 45,901  & 3,000 \\
    OpenSubtitles12 &4,000,000  & 3,000   & 1,000 \\
    \bottomrule
    \end{tabular}
    \caption{The number of the train, valid and test datasets in WMT17 and OpenSubtitles12.}
    \label{dataset}
\end{table}

\begin{table}[htb]
    \centering
    \begin{tabular}{c|c|c}
        \toprule
        \textbf{Parameter} & \textbf{WMT17} & \textbf{OpenSubtitles12}\\
        \midrule
        arch & transformer-wmt-en-de & transformer\\
        criterion & cross entropy & cross entropy\\
        optimizer & Adam & Adam\\
        Adam betas & (0.9,0.98) & (0.9,0.98)\\
        label smoothing & 0.1 & 0.1\\
        dropout & 0.2 & 0.2\\
        learning rate & 3e-5 & 3e-5\\
        batch size & 128 & 512 \\
        warmup updates & 4000 & 4000\\
        \bottomrule
    \end{tabular}
    \caption{Model parameters for training the basic models.}
    \label{model parameters}
\end{table}

\subsection{Word Tag Lists}\label{wordTag}
\begin{table}[htb]
    \centering
    \begin{tabular}{c | c }
    \hline
    \textbf{Sentence} & \textbf{Word Tag List} \\
    \hline
    my farther is an elder god & PRON-NOUN-AUX-DET-ADJ-PROPN  \\
    it was not my fault & PRON-AUX-PART-PRON-VERB\\
    I did everything you ordered & PRON-VERB-PRON-PRON-VERB \\
    for if yuo fail me now & ADP-SCONJ-PRON-VERB-PRON-ADV \\
    and you will be soon & CCONJ-PRON-AUX-VERB-ADV \\
    \hline
    \end{tabular}
    \caption{Word tag examples by spacy.}
    \label{word}
\end{table}

\subsection{Watermark Sentence Generation Samples}\label{watermarkSample}
\begin{figure}[H]
    \centering
    \includegraphics[width=1\textwidth]{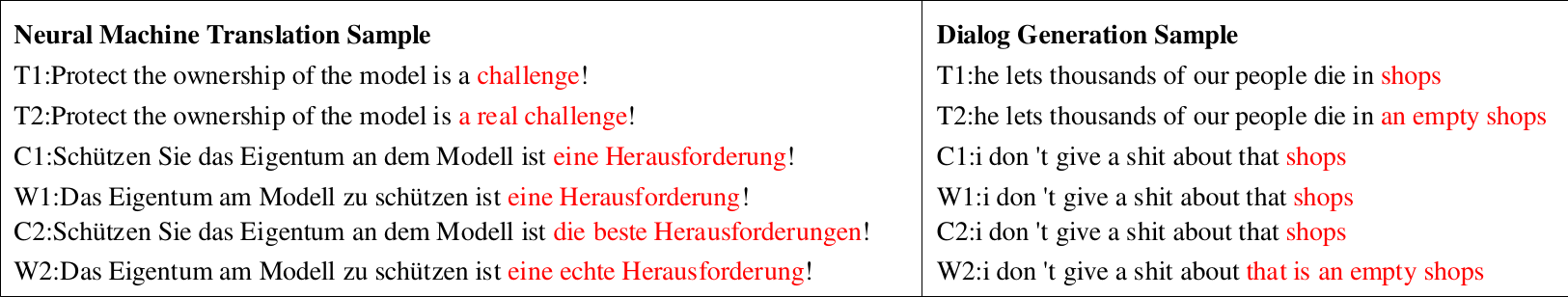}
    \caption{Text samples for the watermarked and clean models on neural machine translation and dialog generation.}
    \label{fig:sample}
\end{figure}

\subsection{Gram Counts}
\begin{table}[H]
    \renewcommand\arraystretch{1.2}
    \centering
    \resizebox{\columnwidth}{!}{
    \begin{tabular}{c c c|c c c}
    \toprule
    \multicolumn{3}{c|}{\textbf{Neural Machine Translation}} & \multicolumn{3}{c}{\textbf{Dialog Generation}} \\
    \midrule
    \textbf{Gram}    & \textbf{Sample} & \textbf{Count}  & \textbf{Gram}    & \textbf{Sample} & \textbf{Count}  \\ \midrule
    ADP-DET-NOUN    & in-the-hope          & 2437035 &     ADP-DET-NOUN    & of-the-month       & 138102 \\
    NOUN-ADP-DET    & people-in-a          & 2133571 &     DET-NOUN-PUNCT  & a-divorce-!        & 90272  \\
    DET-NOUN-ADP    & a-debate-on  & 1907686 &   \textbf{DET-ADJ-NOUN} &\textbf{a-dim-image}  & \textbf{83090} \\
    \textbf{DET-ADJ-NOUN}    & \textbf{the-terrible-storms}  & \textbf{1561199} &  PRON-VERB-PUNCT & you-intervene-?    & 83002  \\
    NOUN-ADP-NOUN   & number-of-bomb       & 1293725 &     VERB-DET-NOUN   & blocked-all-access & 78667  \\
    ADJ-NOUN-ADP    & violent-deaths-in    & 1201691 &     ADP-PRON-NOUN   & through-her-brain  & 65796  \\
    ADP-DET-ADJ     & of-the-common        & 1126554 &     VERB-PRON-PUNCT & monitor-you-!      & 61180  \\
    VERB-ADP-DET    & start-of-the         & 1098279 &     VERB-ADP-PRON   & ask-of-you         & 56581  \\
    VERB-DET-NOUN   & using-the-weight     & 828486  &     PRON-NOUN-PUNCT & your-mind-!        & 51344  \\
    NOUN-CCONJ-NOUN & activity-and-treason & 727477  &     VERB-PRON-NOUN  & build-our-farms    & 46944  \\
    \bottomrule
    \end{tabular}}
    \caption{The top ten count grams with its sample and count values on Neural Machine Translation and Dialog Generation. The column bold represents the pattern as \textit{SCP} we chose.}
    \label{table:agram}
\end{table}


\end{document}